\documentclass[amsmath,10pt,amssymb,twocolumn]{revtex4}

\usepackage{color}
\usepackage[]{graphicx}
\usepackage{latexsym}
\usepackage{natbib}
\usepackage{amsmath}
\usepackage[caption=false]{subfig}
\usepackage{multirow}

\usepackage{listings}
\usepackage{color}

\definecolor{dkgreen}{rgb}{0,0.6,0}
\definecolor{gray}{rgb}{0.5,0.5,0.5}
\definecolor{mauve}{rgb}{0.58,0,0.82}

\usepackage[final]{feynmp}
\newcommand{\s}{\sigma}

\newcommand{\la}{\langle}
\newcommand{\ra}{\rangle}

\newcommand{\be}{\begin{equation}}
\newcommand{\ee}{\end{equation}}
\newcommand{\bc}{\begin{cases}}
\newcommand{\ec}{\end{cases}}
\newcommand{\CI}{\mathrm{CI}}
\newcommand{\xfin}{x^{\infty}}
\newcommand{\xin}{x^0}

\begin{document}

\title{Collective Influence Algorithm to find influencers via
  optimal percolation in massively large social media}

\author{Flaviano Morone}
\author{Byungjoon Min}
\author{Lin Bo}
\author{Romain Mari}
\author{Hern\'an A. Makse}

\affiliation{Levich Institute and Physics Department, City College of
  New York, New York, NY 10031}

\begin{abstract}
We elaborate on a linear time implementation of the Collective
Influence (CI) algorithm introduced by Morone, Makse, {\it Nature}
{\bf 524}, 65 (2015) to find the minimal set of influencers in a
complex network via optimal percolation. We show that the
computational complexity of CI is $O(N\log N)$ when removing nodes
one-by-one, with $N$ the number of nodes in the network. This is made
possible by using an appropriate data structure to process the CI
values, and by the finite radius $\ell$ of the CI sphere.
Furthermore, we introduce a simple extension of CI when
$\ell\to\infty$, the CI propagation ($\CI_{\rm P}$) algorithm, that
considers the global optimization of influence via message passing in
the whole network and identifies a slightly smaller fraction of
influencers than CI. Remarkably, $\CI_{\rm P}$ is able to reproduce
the exact analytical optimal percolation threshold obtained by Bau,
Wormald, {\it Random Struct. Alg.} {\bf 21}, 397 (2002) for cubic
random regular graphs, leaving little improvement left for random
graphs. We also introduce the Collective Immunization Belief
Propagation algorithm ($\CI_{\rm BP}$), a belief-propagation (BP)
variant of CI based on optimal immunization, which has the same
performance as $\CI_{\rm P}$.  However, this small augmented
performance of the order of $1-2 \%$ in the low influencers tail comes
at the expense of increasing the computational complexity from
$O(N\log N)$ to $O(N^2\log N)$, rendering both, $\CI_{\rm P}$ and
$\CI_{\rm BP}$, prohibitive for finding influencers in modern-day
big-data. The same nonlinear running time drawback pertains to a
recently introduced BP-decimation (BPD) algorithm by Mugisha, Zhou,
{\it arXiv:1603.05781}. For instance, we show that for big-data social
networks of typically $200$ million users (eg, active Twitter users
sending 500 million tweets per day), CI finds the influencers in less
than 3 hours running on a single CPU, while the BP algorithms
($\CI_{\rm P}$, $\CI_{\rm BP}$ and BDP) would take more than 3,000
years to accomplish the same task.

\end{abstract}

\maketitle

In Ref. \cite{MoMa} we developed the theory of influence maximization
in complex networks, and we introduced the Collective Influence (CI)
algorithm for localizing the minimal number of influential nodes.  The
CI algorithm can be applied to a broad class of problems, including
the optimal immunization of human contact networks and the optimal
spreading of informations in social media, which are ubiquitous in
network science. In fact, these two problems can be treated in a
unified framework. As we noticed in~\cite{MoMa}, the concept of
influence is tightly related to the concept of network integrity. More
precisely, the most influential nodes in a complex network form the
minimal set whose removal would dismantle the network in many
disconnected and non-extensive components.  The measure of this
fragmentation is the size of the largest cluster of nodes, called the
giant component $G$ of the network and the problem to find the minimal
set of influencers can be mapped to optimal percolation.


The influence maximization problem is NP-hard, and it can be
approximately solved by different methods. We showed in \cite{MoMa}
that the objective function of this optimization problem is the
largest eigenvalue of the Non-Backtracking matrix (NB) of the network
$\lambda_{\rm max}(\vec{n})$, where $\vec{n}=(n_1,n_2\dots, n_N)$ is
the vector of occupation numbers encoding node's vacancy ($n_i=0$) or
occupancy ($n_i=1$).  In \cite{MoMa} we introduced the Collective
Influence algorithm to minimize $\lambda_{\rm max}(\vec{n})$. This
algorithm is able to produce nearly optimal solutions in almost linear
time, and performs better than any other algorithm with comparable,
i.e. nearly linear, computational running time.

In this paper we describe an improved implementation of the original
CI algorithm, which keeps the computational complexity bounded by
$O(N\log N)$ even when nodes are removed one-by-one. This is made
possible by the finite size of the Collective Influence sphere, which,
in turn, allows one to use a max-heap data structure to process very
efficiently the CI values.  The linear time implementation of CI is
explained in Section \ref{sec:CImaxheap}.

\begin{figure}[h!]
\includegraphics[width=.5\textwidth]{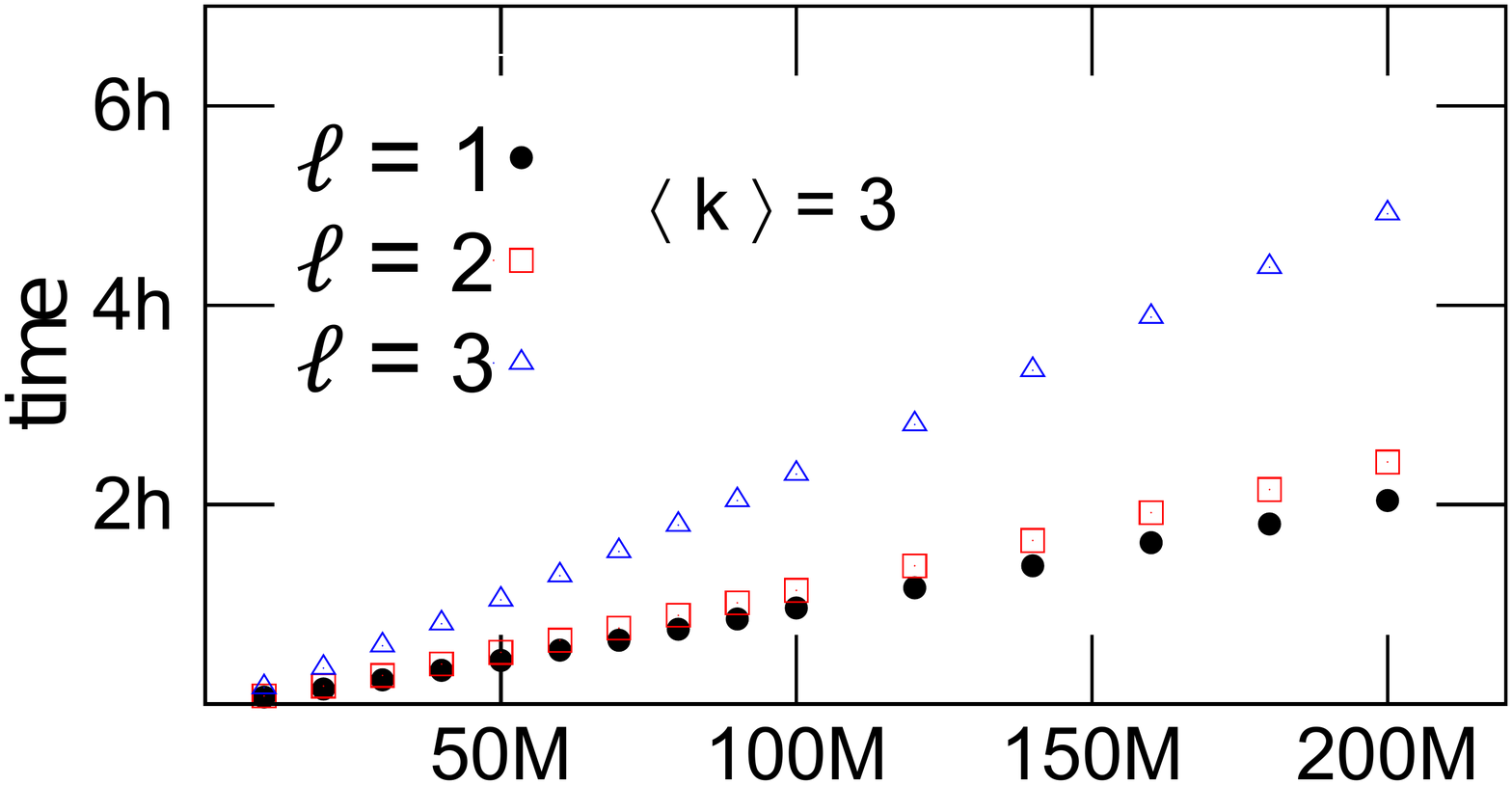}
\caption{Running time of the CI algorithm (including the reinsertion
  step) for ER random graphs of average degree $\la k\ra=3$, as a
  function of the network size, and for different values of the radius
  $\ell$ of the ball.  (To generate very large ER random graphs we
  used the algorithm of Ref. \cite{fast_generator}).  For a graph with
  $0.2$ billion nodes the running time is less than $2.5$ hours with
  $\ell=2$ and 5 hours with $\ell=3$.}
\label{fig:runtime-CI}
\end{figure}

In Section \ref{sec:CIpro} we introduce a generalized version of the
CI algorithm, which we name Collective Influence Propagation
($\CI_{\rm P}$), that incorporates the information about nodes
influence at the global level. Indeed, it can be seen as the limit
version of CI when the radius $\ell$ of the ball is sent to
infinity. The $\CI_{\rm P}$ algorithm allows one to obtain slightly
better solutions to the problem, i.e., a smaller set of optimal
influencers than CI. Remarkably, it is able to reach the exact optimal
percolation threshold in random cubic graphs, as found analytically by
Bau et al.~\cite{wormald}.  However, this augmented performance comes
at the expense of increasing the computational complexity of the
algorithm from $O(N\log N)$ to $O(N^2\log N)$. The same nearly
quadratic running time pertains also to a
Belief-Propagation-Decimation (BPD) algorithm recently suggested by
Mugisha and Zhou in Ref.~\cite{zuo}, as we show in
Fig.~\ref{fig:runtime}.  Based on this observation, CI remains the
viable option for a fast and nearly-optimal influencer search engine
in massively large networks. Quantitatively, a network of $200$
millions nodes can be fully processed by CI (using a radius $\ell=2$)
in roughly 2.5 hours, while both $\CI_{\rm P}$ and BPD would take a
time of the order of $3,000$ years to accomplish the task, as we show
in Figs.~\ref{fig:runtime-CI} and \ref{fig:runtime}.

In Section \ref{sec:collective_immunization} we present yet another
algorithm to solve the optimal influence problem, that we name
Collective Immunization (CIm). The CIm algorithm is a
belief-propagation-like algorithm, which is inspired by the SIR
disease spreading model, and it also gives nearly optimal solutions,
as seen in Fig.~\ref{fig:rrg_warmald}.






\section{Implementing CI in linear time.}
\label{sec:CImaxheap}

In this section we describe how to implement the CI algorithm 
to keep the running time $O(N\log N)$ even when the nodes are 
removed one-by-one.

CI is an adaptive algorithm which removes nodes progressively
according to their current CI value, given by the following formula: 
\be
\CI_{\ell}(i)\ =\ (k_i-1)\sum_{j\in\partial B(i,\ell)}(k_j-1)\ , 
\label{eq:CIformula}
\ee At each step, the algorithm removes the node with the highest
$\CI_{\ell}(i)$ value, and keep doing so until the giant component is
destroyed.  A straightforward implementation of the algorithm consists
in computing at each step the $\CI_{\ell}(i)$ for each node $i$, sort
these values, and then removing the node with the largest $\CI_{\ell}$
value.  Despite its simplicity, this implementation is not optimal, as
it takes a number of operations of the order $O(N^2\log N)$.

However, the time complexity of the CI-algorithm can be kept at
$O(N\log N)$ by using an appropriate data structure for storing and
processing the CI values. The basic idea is that, after each node
removal, we would like to recompute the CI of a $O(1)$ number of other
nodes and we would like to avoid sorting and sorting again after each
update, since we only need the largest CI value at each step, and thus
is useless to have a completely sorted list of values. This idea can
be realized by using a max-heap data structure.
 
\begin{figure}[h!]
\includegraphics[width=.5\textwidth]{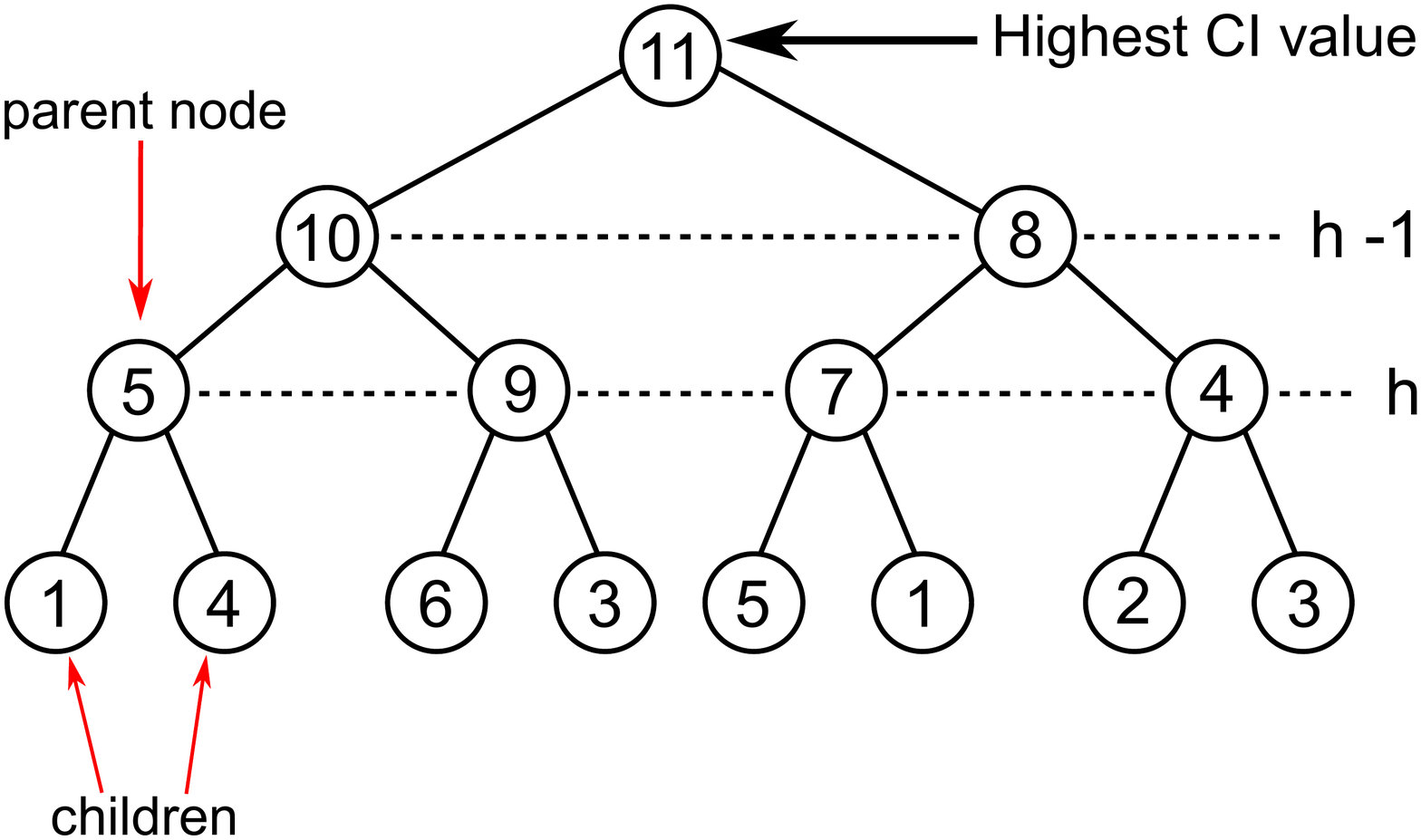}
\caption{Max heap data structure used to implement the CI algorithm. In the max heap 
each parent node stores a CI value larger than the ones stored by its children. 
No ordering prescription is imposed to the nodes belonging to the same level
$h$ of the heap.}
\label{fig:heap}
\end{figure}

Before to delve into the details, let us recall the definition of a
"heap".  A heap is a binary tree encoding a prescribed hierarchical
rule between the parent node at level $h$ and its children nodes at
level $h+1$, with no hierarchy among the children.  In our specific
case we use a heap with a max heap rule, i.e., each parent node of the
heap stores a CI value greater or equal to those of the children, but
there is no order between the left child and right one (see
Fig. \ref{fig:heap}). The root node of the max heap stores
automatically the largest CI value.


One more concept is needed, i.e., the concept of "heapification", which we shall be using 
often later on.
Generally speaking, given a set of numbers $S=\{x_1,\dots,x_N\}$, the 
heapification of the set $S$ is a permutation $\Pi$ of the elements 
$\{x_{\Pi(1)},\dots,x_{\Pi(N)}\}$ satisfying the following max-heap property:
\be
x_{\Pi(i)}\geq x_{\Pi(2i)}\ \ \and \text{AND}\ \ x_{\Pi(i)}\geq x_{\Pi(2i+1)}\ .
\ee

We call heapify($i$) the function which heapifies the CI values in the sub-tree 
rooted on node $i$. The aim of this function is to down-move node $i$ in the heap 
by swapping it with the largest of its children until it satisfies the max-heap property 
in the final location.

%

\begin{figure}[h!]
\includegraphics[width=.4\textwidth]{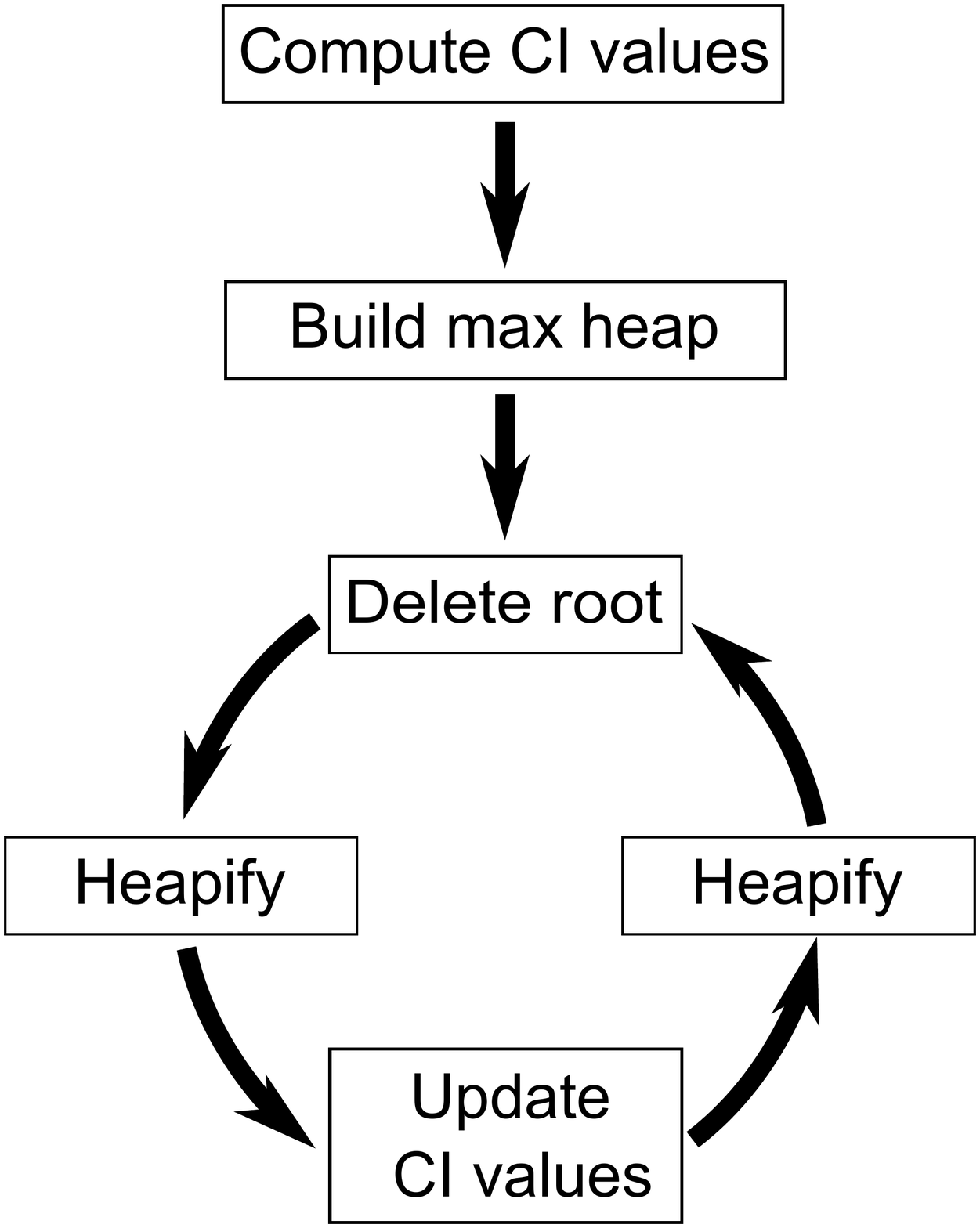}
\caption{Flow of the CI algorithm. The first part of the algorithm,
  executed only once, consists of two steps: i) computing CI for each
  node, and ii) allocating the CI values in the max-heap. After that,
  it follows the main loop of the algorithm, which consists of three
  steps: iii) removing the node with highest CI value along with the
  root of the heap; iv) heapifying the heap starting from the new root
  (see Step3); v) updating the CI values of the perturbed nodes, and
  heapifying the sub-trees rooted on each updated node.  The loop ends
  when the giant component is destroyed.}
\label{fig:flow}
\end{figure}

Having defined the main tools we are going to use in the implementation, 
we can now discuss the flow of the algorithm step by step, as schematized in
Fig. \ref{fig:flow}

\medskip

{\bf Step 1 - Computing CI}. To compute the $\CI_{\ell}(i)$ value of
node $i$ according to Eq. \eqref{eq:CIformula} we have to find the
nodes belonging to the frontier $\partial B(i,\ell)$ of the ball of
radius $\ell$ centered on $i$ (we define the distance between two
nodes as the number of edges of the shortest path connecting them).
In an undirected network the nodes $j\in \partial B(i,\ell)$ can be
found using a simple breadth-first-search (BFS) up to a distance
$\ell$ from the central node $i$.  First we visit the nearest
neighbours of node $i$, which, of course, belong to $\partial
B(i,1)$. Then we visit all the neighbours of those nodes not yet
visited, thus arriving to $\partial B(i,2)$.  We keep on going until
we visit all the nodes in $\partial B(i,\ell)$. At this point we use
the nodes $j\in \partial B(i,\ell)$ to evaluate $\CI_{\ell}(i)$ using
Eq. \eqref{eq:CIformula}.

When all the CI values $\{\CI_{\ell}(1),\dots,\CI_{\ell}(N)\}$ have
been calculated, we arrange them in a max heap, as explained next.

\medskip

{\bf Step2 - Building the max-heap}.
We build the heap in a bottom-up fashion, from the leaves to the root.
Practically, we first fill the heap with arbitrary values and then we heapify 
all the levels starting from the lowest one. In this way the root stores 
automatically the largest CI value.

\medskip

{\bf Step3 - Removal}.
We remove from the network the node having the largest CI value, and 
we decrement by one the degrees of its neighbors. The largest CI value is 
stored in the root of the max-heap. Therefore, after the removal, the 
root in the max heap has to be replaced by the new largest CI value.
The easiest way to do this is replacing the root with the rightmost leaf in the 
last level of the heap, decreasing the size of the heap by one, and heapifying the  
new root.

\begin{figure}[h!]
\includegraphics[width=.5\textwidth]{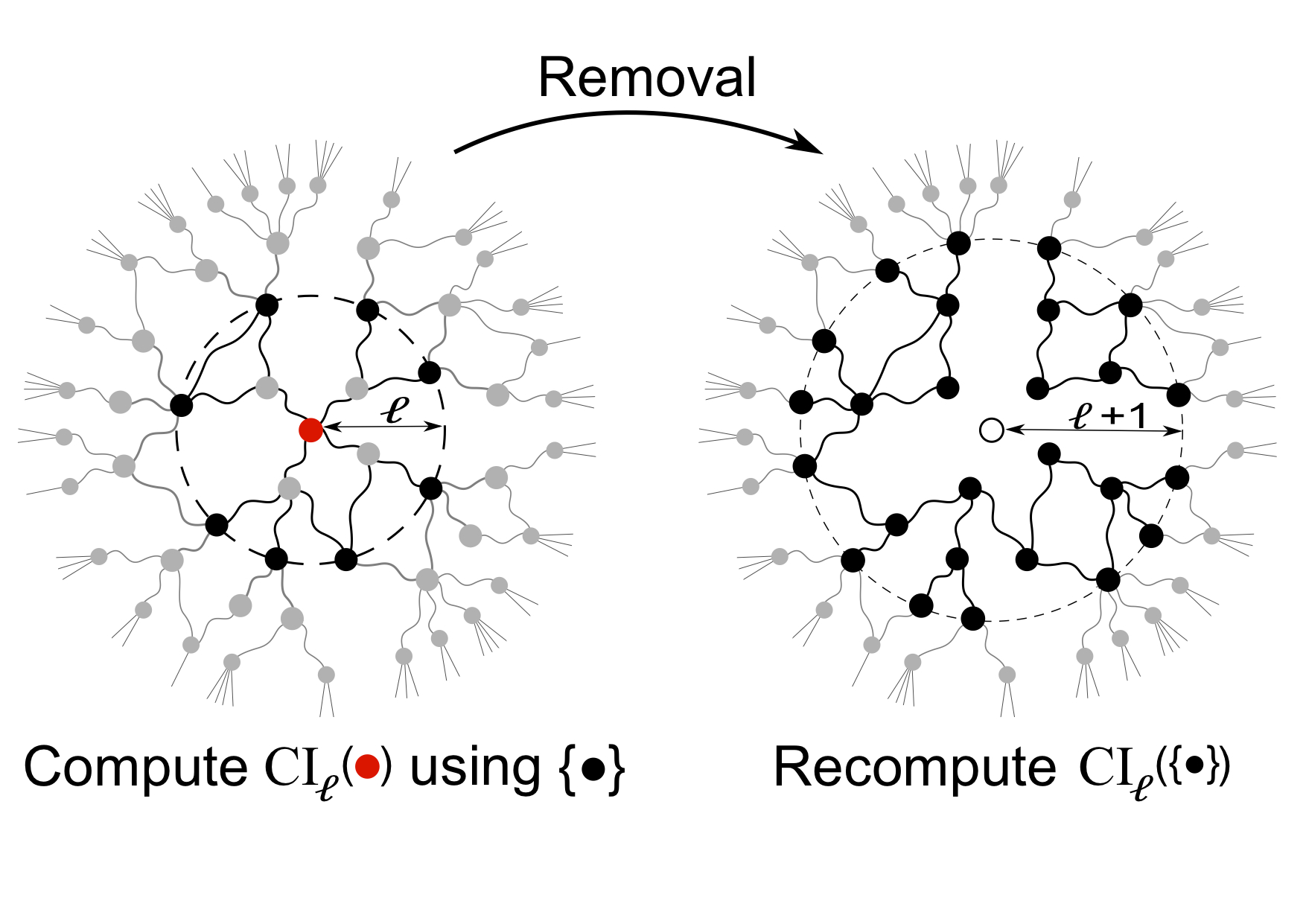}
\caption{ {\bf Left panel}: the CI of the red node at the level $\ell$
  is computed using the nodes on the boundary of the ball of radius
  $\ell$ centered on the red node.
{\bf Right panel}: the removal of the red node perturbs the CI values
of nodes located up to a distance $\ell + 1$ from it. Accordingly,
only the CI values of these nodes (the black ones) have to be updated
before the next removal.}
\label{fig:relevantNodes}
\end{figure}

\medskip
{\bf Step4 - Updating CI values}.
The removal of a node perturbs the CI values of other nodes, that must
be recomputed before the next removal. The nodes perturbed by the
removal are only the ones placed at distances $1,2,\dots,\ell,\ell+1$
from the removed one.  In other words, only the nodes inside the ball
$\mathrm{B}(i,\ell+1)$ change their CI values when $i$ is removed,
while the others remain the same (see Fig. \ref{fig:relevantNodes}).

The CI values of nodes on the farthest layer at $\ell+1$ are easy to
recompute.  Indeed, let us consider one of this node and let us call
$k$ its degree.  After the removal of the central node its CI value
decreases simply by the amount $k-1$.  For nodes in the other layers
at distance $1,2,\dots,\ell$, the shift of their CI values is, in
general, not simple to assess, and we need to use the procedure
explained in Step1.

When we modify the CI value stored in a node of the heap, it may
happen that the new heap does not satisfy the max-heap rule. Therefore
we have to restore the max heap-structure after each change of the CI
values.  More precisely, we proceed as follows. Let us consider one
among the nodes to update.  Assuming that the structure around the
removed node is locally tree-like, the new CI values of the
surrounding nodes can only be smaller than the old ones, and,
consequently, we need to heapify only the sub-tree rooted on those
nodes.  We stress that the order of the update-heapification
operations is important: each node update must be followed by the
corresponding heapification, before updating the next node.


\medskip

\subsection{Running time}

The running time of the CI algorithm is $O(N\log N)$.  In fact, Step1
and Step2 take both $O(N)$ operations and they are performed only
once.  Step3 and Step4 take each at most $O(\log N)$ operations and
they are repeated $O(N)$ times. Therefore the algorithm takes $O(N\log
N + N)\sim O(N\log N)$ operations.  To check the $N\log N$ scaling of
the CI algorithm we performed extensive numerical simulations on very
large networks up to $N = 2\times 10^8$ nodes. The results shown in
Fig. \ref{fig:runtime-CI} clearly confirm that the CI algorithm runs in
nearly linear time.

\medskip

\begin{figure}[h!]
\includegraphics[width=.5\textwidth]{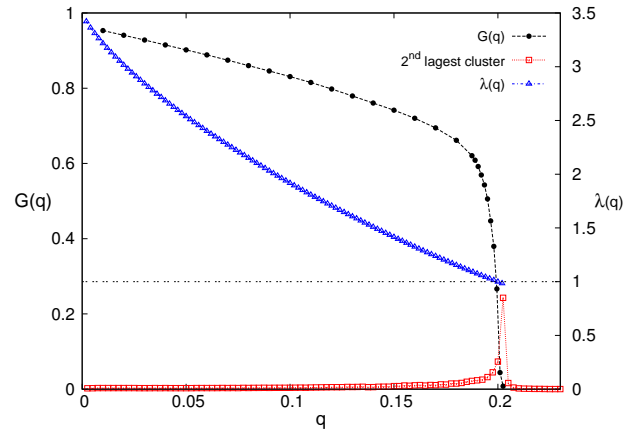}
\caption{Giant component $G(q)$ (black dots) computed with CI, second
  largest cluster (red squares), and the eigenvalue $\lambda(\ell; q)$
  Eq. \eqref{eq:lambda_CI}, as a function of the removed nodes
  $q$. Here we used an ER network of $10^6$ nodes, average degree $\la
  k\ra = 3.5$, and a value $\ell=5$ for CI algorithm.  The eigenvalue
  $\lambda(q)$ reaches one when the giant component is zero, as
  signaled also by the peak in the size of the second largest
  cluster. The size of the second largest cluster is magnified to make
  it visible at the scale of the giant component.  }
\label{fig:lambda} 
\end{figure}

\begin{figure}[h!]
\includegraphics[width=.5\textwidth]{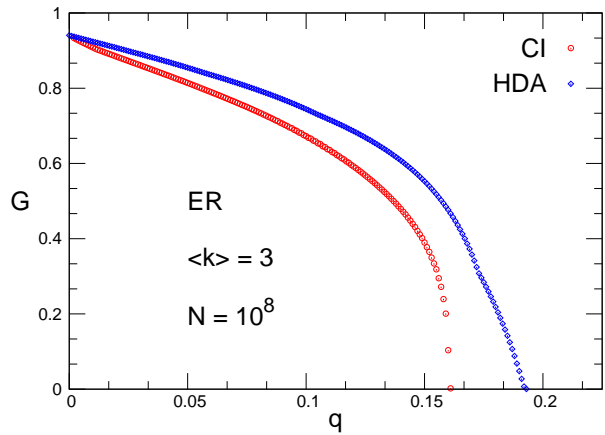}
\caption{CI algorithm applied to a ER network of $10^8$ nodes (red
  circles).  The result is compared with the HDA (high degree
  adaptive) strategy (blue diamonds); one of the few strategies which
  is adaptive and linear in algorithmic time. Indeed, the same
  max-heap idea we used here in CI can be used in other types of
  adaptive algorithms which share the same properties as CI. For
  example it is easy to see that the adaptive high degree strategy can
  be implemented in the same way, and therefore has the same running
  time as the non adaptive high degree attack.}
\label{fig:bigNet}
\end{figure}

{\bf Step5 - Stopping the algorithm} To decide when the algorithm has
to be terminated we use a very simple method, which allows one to
avoid checking when the giant component $G$ vanishes.  The idea is to
monitor the following quantity after each node removal: \be
\begin{aligned}
\lambda(\ell; q)\ &=\ \left(\frac{\sum_i\CI_{\ell}(i) }{N\la k\ra}\right)^{1/(\ell+1)}, \\
\end{aligned}
\label{eq:lambda_CI}
\ee where $\la k\ra$ is the average degree of the network for $q=0$.
Equation \eqref{eq:lambda_CI} gives an approximation of the minimum of
the largest eigenvalue of the non-backtracking matrix when $Nq$ nodes
are removed from the network~\cite{MoMa}.  For $q=0$, it is easy to
show that, for tree-like random graphs, $\lambda(\ell; 0) = \kappa -
1$, where $\kappa = \la k^2\ra/\la k\ra$.  Removing nodes decreases
the eigenvalue $\lambda(\ell; q)$, and the network is destroyed when
$\lim_{\ell\to\infty}\lambda(\ell; q=q_c)=1$.  Practically we cannot
take the limit $\ell\to \infty$, but for a reasonably large $\ell$,
the relaxed condition $\lambda(\ell; q=q_c)=1$ works pretty well, as
we show in Fig. \ref{fig:lambda}.  Therefore, we can stop the
algorithm when $\lambda(\ell; q)=1$.  The advantage of
Eq. \eqref{eq:lambda_CI} is that it can be updated on runtime at
nearly no additional computational cost, and therefore does not
require additional $O(N)$ calculations needed to compute the giant
component.  Figure \ref{fig:bigNet} shows the giant component attacked
by CI and high-degree adaptive in a ER network of 100 million nodes.

\subsection{Reinsertion}
We conclude this section by discussing a refinement of CI algorithm,
which we use to minimize the giant component in the phase $G>0$. This
can be useful when it is not possible to reach the percolation
threshold (where $G=0$), but one still wants to minimize $G$ using the
available resources, i.e., the maximum number of node removals at
one's disposal.  The main idea is based on a reinsertion method,
according to which nodes are reinserted in the network using the
following criterion.  We start from the percolation point, where the
network is fragmented in many clusters. We add back in the network one
of the removed node, which is chosen such that, once reinserted, it
joins the smallest number of clusters. Note that we do not require
that the reinserted node joins the clusters of smallest sizes, but
only the minimum number of clusters, independently from their
sizes. When the node is reinserted we restore also the edges with its
neighbors which are in the network (but not the ones with neighbors
not yet reinserted, if any). The procedure is repeated until all the
nodes are back in the network. When implementing the reinsertion, we
add back a finite fraction of nodes at each step. In our simulations
we reinserted $0.2\%$ of nodes at each step. Moreover we observed that
even using a smaller fraction than $0.2\%$, we obtained the same
results.

\section{CI propagation}
\label{sec:CIpro}
In this section we present the CI-propagation algorithm ($\CI_{\rm
  P}$), which extends the CI algorithm to take into account the global
information beyond the local CI sphere. However, the main idea of
$\CI_{\rm P}$ remains the same, i.e., minimizing the largest eigenvalue of
the Non-Backtracking (NB) matrix~\cite{MoMa}. Indeed, $\CI_{\rm P}$ is
obtained asymptotically from $\CI_\ell$ as $\ell\to \infty$.
 
The NB is a non-symmetric matrix and it has different right and left
eigenvectors.  As we will see the right and left eigenvectors
corresponding to the largest eigenvalue provides two different, yet
intuitive, notions of node's influence.  The left eigenvector
$\vec{L}$ is a vector with $2M$ entries $L_{i\to j}$, where $M$ is the
total number of links, that satisfies the following equation: \be
L_{i\to j} = \frac{1}{\lambda_{\rm max}}\ \sum_{k\in \partial
  i\setminus j}L_{k\to i}\equiv\ \frac{1}{\lambda_{\rm
    max}}(\vec{L}\hat{\mathcal{B}}^T)_{i\to j}\ .
\label{eq:left}
\ee
A similar equation holds for the right eigenvector $\vec{R}$:
\be
R_{i\to j} = \frac{1}{\lambda_{\rm max}}\ 
\sum_{k\in \partial j\setminus i}R_{j\to k}\equiv\ 
\frac{1}{\lambda_{\rm max}}(\hat{\mathcal{B}}\vec{R})_{i\to j}\ , 
\label{eq:right}
\ee
where $\hat{\mathcal{B}}$ is the NB matrix.
Both left and right eigenvectors can be thought of as two set of 
messages traveling along the directed edges of the network. 
This becomes more apparent if we transform Eqs. \eqref{eq:left}-\eqref{eq:right} 
in dynamical updating rules for the messages $L_{i\to j}$ and $R_{i\to j}$ as:
\be
\begin{aligned}
L_{i\to j}^t &= \frac{1}{\lambda_{\rm max}^{t-1}}\ 
\sum_{k\in \partial i\setminus j}L_{k\to i}^{t-1}\equiv\ 
\frac{1}{\lambda_{\rm max}^{t-1}}(\vec{L}\mathcal{B}^T)_{i\to j}\ ,\\
R_{i\to j}^t &= \frac{1}{\lambda_{\rm max}^{t-1}}\ 
\sum_{k\in \partial j\setminus i}R_{j\to k}^{t-1}\equiv\ 
\frac{1}{\lambda_{\rm max}^{t-1}}(\mathcal{B}\vec{R})_{i\to j}\ , \\
\lambda_{\rm max}^{t-1} &=\ 
\sqrt{\sum_{{\rm All}\ i\to j}(L_{i\to j}^{t-1})^2} = 
\sqrt{\sum_{{\rm All}\ i\to j}(R_{i\to j}^{t-1})^2}\ .
\label{eq:dyn_messages}
\end{aligned}
\ee
The interpretation of Eqs. \eqref{eq:dyn_messages} is the following.
For each directed edge $i\to j$, the message $L_{i\to j}^t$ at time $t$ 
from $i$ to $j$ is updated using the messages $L_{k\to i}^{t-1}$ incoming 
into node $i$ at time $t-1$, except the message $L_{j\to i}^{t-1}$. 
Therefore, the left message $L_{i\to j}^t$ represents the amount of information 
 {\bf received} by node $i$ from its neighbours, other than $j$. 
On the contrary, the right message $R_{i\to j}$ is updated using the sum of the 
outgoing messages from node $j$ to nodes $k$ other than $i$, and thus it 
measures the amount of information {\bf sent} by node $j$ to its 
neighbours, other than $i$.

Now we come to the problem of minimizing $\lambda_{\rm max}$ by
removing nodes one-by-one. According to the discussion above, we can
measure the influence of each node in different ways. The easiest one
is to assign to each node $i$ the sum of all the incoming left
messages $L_{k\to i}$: \be \CI_{\rm IN}(i)\ = \sum_{k\in\partial
  i}L_{k\to i}\ .
\label{eq:in-fluence}
\ee
The interpretation of this quantity comes directly from the recursive Eq. 
\eqref{eq:left}. Indeed, if we plug into \eqref{eq:in-fluence} the recursion 
for $L_{k\to i}$ given by \eqref{eq:left}, and we keep on iterating $\ell$ times, 
we see that the influence of node $i$ is determined by the sum of all the  
messages $L_{\to \rm{Ball}(i,\ell)}$ incoming into the ball of radius $\ell$ 
centered on $i$, which has an evident similarity with the usual CI definition. 

Another possibility is to assign to node $i$ the sum of all the incoming right 
messages $R_{k\to i}$:
\be
\CI_{\rm OUT}(i)\ = \sum_{k\in\partial i}R_{k\to i}\ .
\label{eq:out-fluence}
\ee This quantity is the dual of the previous one, and therefore we
used the name $\CI_{\rm OUT}$. Indeed, by proceeding as before, i.e.,
plugging Eq. \eqref{eq:right} into \eqref{eq:out-fluence} and
iterating $\ell$ times, we see that the influence of node $i$ is now
determined by the sum of all the messages $R_{\rm{Ball}(i,\ell)\to}$
outgoing from the ball of radius $\ell$ centered on $i$, which again
bears a close similarity with CI.  We could say that
Eq. \eqref{eq:in-fluence} measures the "IN-fluence" of node $i$, while
Eq. \eqref{eq:out-fluence} measures its "OUT-fluence". Since we
believe that both measures do capture a specific aspect of the
importance of a given node, we combine them in what we call the
Collective Influence Propagation, which is defined as: \be \CI_{\rm
  P}(i)\ =\ \sum_{k\in\partial i}\sqrt{L_{k\to i}R_{k\to i}}\ .
\label{eq:cipro}
\ee

The quantity $\CI_{\rm P}(i)$ combines both the information received
and the information propagated by node $i$.

Having defined the main quantity of the $\CI_{\rm P}$ algorithm, we
move to explain the few simple steps to implement it.
\begin{itemize}
\item 1) Start with all nodes present and iterate
  Eqs. \eqref{eq:dyn_messages} until convergence.
\item 2) Use the converged messages $L_{i\to j}$ and $R_{i\to j}$ to compute the 
$\CI_{\rm P}(i)$ values for each node $i$.
\item 3) Remove node $i^*$ with the highest value of $\CI_{\rm P}(i^*)$ and set to zero 
all its ingoing and outgoing messages. 
\item 4) Repeat from 2) until $\lambda_{\rm max}=1$.
\end{itemize}

\begin{figure}[h!]
\includegraphics[width=.5\textwidth]{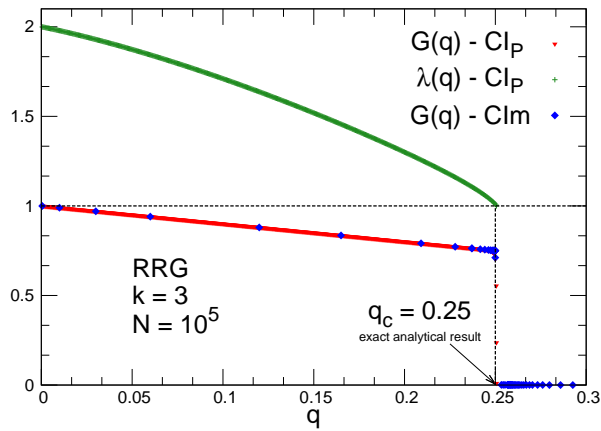}
\caption{Giant components $G(q)$ (red triangles and blue diamonds) 
computed with the $\CI_{\rm P}$ and the CIm algorithms, 
and the eigenvalue $\lambda(q)$ (green crosses) computed with $\CI_{\rm P}$, 
as a function of the removed nodes $q$, in a Random Regular Graph of 
$10^5$ nodes, and degree $ k = 3$. The vertical line at 
$q = 0.25 = q_c$ marks the position of the analytical 
exact optimal value of the percolation threshold~\cite{wormald}.
}
\label{fig:rrg_warmald}
\end{figure}

\begin{figure}[h!]
\includegraphics[width=.5\textwidth]{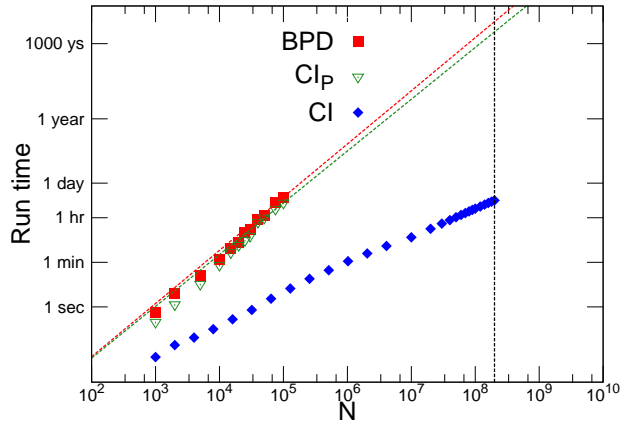}
\caption{Running time of CI (blue diamonds) at level $\ell=3$,
  $\CI_{\rm P}$ (green triangles), and the BPD algorithm of
  Ref.~\cite{zuo}, as a function of the network size $N$, for ER
  networks of average degree $\la k \ra = 3$. The CI algorithm is the
  only one that scales almost linearly with the system size, while
  both BP algorithms, $\CI_{\rm P}$ and BPD, scale quadratically with
  the network size $N$.  The vertical dashed line is at $N = 2 \times
  10^8$: for this network size, the running time of CI at level
  $\ell=3$ is roughly $5$ hours and $2.5$ hours for $\ell=2$, while
  both $\CI_{\rm P}$ and BPD would take a time of $\sim 3,000$ years
  to accomplish the same task.  (To measure the running time of
  $\CI_{\rm P}$ and BPD we used the same number of iterations of the
  messages.  Data are in $\log-\log$ scale.)  }
\label{fig:runtime}
\end{figure}

The $\CI_{\rm P}$ algorithm produces better results than CI. As we
show in Fig. \ref{fig:rrg_warmald} for the case of a random cubic
graph, $\CI_{\rm P}$ is able to identify the optimal fraction of
influencers, which is known analytically to be $q_c=1/4$ \cite{wormald}.
Unfortunately the $\CI_{\rm P}$ algorithm has running time $O(N^2\log
N)$ and thus cannot be scaled to very large networks, as we show in
Fig.~\ref{fig:runtime}, where we also compare with the time complexity
of the BPD algorithm of Mugisha, Zhou \cite{zuo} and with the original
CI algorithm.

We close this section by noticing that $\CI_{\rm P}$ is a
parameter-free algorithm, i.e., it does not require any fine tuning
and can be applied straight away due to its low programming
complexity.  The introduction of more parameters (like the
temperature) may still improve the performance of the algorithm. While
it may be an interesting technical problem, we did not develop further
the $\CI_{\rm P}$ algorithm, mainly because the introduction of
external parameters would not reduce anyway the quadratic running
time. Also the quasi-optimal performance of $\CI_{\rm P}$ for finding
minimal percolation sets in small systems in
Fig. \ref{fig:rrg_warmald} leaves little improvement left, so that we
do not develop the algorithm further.

\section{Collective Immunization }
\label{sec:collective_immunization}

In this section we formulate the optimal percolation problem as the
limit of the optimal immunization problem in the SIR
--Susceptible-Infected-Recovered-- disease spreading
model~\cite{kermack}, and we present the Collective Immunization (CIm)
algorithm or $\CI_{\rm BP}$ based on Belief Propagation.

According to the SIR model, a variable $x_i^t = \{ S, I, R \}$ encodes
the state of each node $i$ at time step $t$.  A node in a state
$x_i=I$ stays infected for a finite time, and in this state, it may
infect a neighboring node $j$ if $x_j=S$.  After the infectious
period, the infected node $i$ recovers. Nodes in state $R$ stay in $R$
forever, being immune to further infection. Thus in the long time
limit, the disease state $\xfin_i$ of any node $i$ is either $R$ or
$S$.  In this limit one can compute the marginals of $\xfin$ on any
node, knowing the initial state $\xin$, in a `message passing' manner.
The message that node $i$ passes to node $j$ is the probability
$\nu_{i\to j}(\xfin_i| \xin_i)$ that node $i$ ends in state $\xfin_i$
knowing it starts in state $\xin_i$, assuming that node $j$ is absent.

According to the dynamic rule of SIR model, we have the following set
of relations:
\begin{equation}
  \begin{array}{ll}
    \nu_{i\to j}(\xfin_i=R| \xin_i=S) & =  
    1 - \nu_{i\to j}(\xfin_i=S|\xin_i=S) \\
    \nu_{i\to j}(\xfin_i=S| \xin_i=R) & =  
    0  \\
    \nu_{i\to j}(\xfin_i=S| \xin_i=I \,\,) & =  
    0  \\
  \end{array}
\end{equation}

Therefore, it is clear that the knowledge of the sole $\nu_{i\to
  j}(\xfin_i=S|\xin_i=S)$ is enough to reconstruct the long time limit
of the marginal of $\xfin_i$. Next, we assume that each node is
initially infected with probability $\gamma$, i.e., at time $0$ a
randomly chosen set of $\gamma N$ sites are infected. We also
introduce a binary variable $n_i$ for each node $i$, taking values
$n_i=0$ if node $i$ is immunized (i.e. removed in the language of
optimal percolation), and $n_i=1$ if it is not (i.e. present).  For a
locally tree-like interaction network (and when the clustering
property holds), the probabilities (messages) received by node $i$
from its neighbors $j$ can be considered as uncorrelated. This allows
one to calculate self-consistently the messages through the following
equations:
\begin{equation}
\label{eq:Pto_immunization}
\begin{aligned}
   & \nu_{i\to j}(\xfin_i=S|\xin_i \neq R) =\\
&= (1-\gamma) \prod_{k\in \partial i\setminus j} 
 \left[ 1 -\beta \left( 1-\nu_{k\to i}(\xfin_k=S|\xin_k \neq R) \right)n_k
      \right],
\end{aligned}
\end{equation}
\noindent where $\beta$ is the transmission probability of the disease
(or the spreading rate). In the end, we will be mainly interested in
the limits $\gamma=1/N$ and $\beta \to 1$.

The marginal probability that node $i$ is eventually susceptible given
that node $i$ is not one of the immunizators is obtained through:
\begin{equation}
\begin{aligned}
  \label{eq:marginal_immunization}
  &\nu_{i}(\xfin_i=S|\xin_i \neq R) =\\ 
 &(1-\gamma) \prod_{k\in \partial i} 
 \left[ 1 - \beta\left( 1-\nu_{k\to i}(\xfin_k=S|\xin_k \neq R) \right)n_k
 \right].
\end{aligned}
\end{equation}
From now on we drop the argument in the probabilities $\nu_{i\to j}$
and $\nu_i$, and we simply write $\nu_{i}(\xfin_i=S|\xin_i \neq R) =
\nu_i$.

The best immunization problem amounts to find the minimal set of
initially immunized nodes that minimizes the outbreak size $F = \sum_i
n_i \left(1- \nu_{i}\right)$.  This problem can be equivalently solved
by minimizing the following energy (or cost) function:
 \begin{equation}
  \label{eq:energy_immunization}
    E(\vec{n}) =  - \sum_i n_i \log \left[ \nu_{i}(\xfin_i=S|\xin_i \neq R) \right].
\end{equation}

The energy function in Eq. \eqref{eq:energy_immunization} has the
virtue of describing a pairwise model, and therefore is easier to
treat. Indeed, substituting~\eqref{eq:marginal_immunization}
into~\eqref{eq:energy_immunization} one can rewrite the energy
function as:
\begin{equation}
  \label{eq:energy_pair_immunization}
\begin{aligned}
   E(\vec{n})  &=  \sum_{<ij>} U_{ij}(n_i, n_j)\ ,\\
  U_{ij}(n_i, n_j)  =
    &-n_i\log \left[ 1 - \beta \left( 1-\nu_{j\to i}\right)n_j
      \right]\\
    &  -n_j\log \left[ 1 - \beta \left( 1-\nu_{i\to j}\right)n_i \right]\ ,
  \end{aligned}
\end{equation}
where we drop an useless constant term.  We found useful to make the
following change of variables:
\begin{equation}
n_i\ =\ \frac{1-\sigma_i}{2}\ ,
\end{equation} 
so that $\s_i = 1$ means that node $i$ is removed or immunized, and
$\s_i=-1$ that it is present.

The minimum of the energy function \eqref{eq:energy_pair_immunization}
can be found by solving the following equations:
\begin{equation}
\begin{aligned}
  \label{eq:cavity_fields_zeroT}
  h^{i} = -\mu + \sum_{k\in \partial i} &\left\{ \max_{\sigma_k}
    \left[- U_{ik}(+1,\sigma_k) + \frac{h^{k\to i}}{2} \sigma_k \right]\right.\\ 
&\left. - \max_{\sigma_k}
    \left[-U_{ik}(-1,\sigma_k) + \frac{h^{k\to i}}{2} \sigma_k \right] 
    \right\}\ ,
\end{aligned}
\end{equation}
\begin{equation}
\begin{aligned}
  \label{eq:cavity_messages_zeroT}
  h_{i\to j} = -\mu + \sum_{k\in \partial i\setminus j} &\left\{ \max_{\sigma_k}
   \left [- U_{ik}(+1,\sigma_k) + \frac{h_{k\to i}}{2}  \sigma_k \right]\right. \\
    &\left.- \max_{\sigma_k}
   \left[-U_{ik}(-1,\sigma_k) + \frac{ h_{k\to i}}{2} \sigma_k \right] 
    \right\}\ ,
\end{aligned}
\end{equation}
where the variable $h_{i}$ is the log-likelihood ratio: 
\begin{equation}
h_{i} = \log\left(\frac{ {\rm probability\ that\ } i {\rm\ is\ removed} }
{{\rm probability\ that\ } i {\rm\ is\ present} }\right)\ ,
\end{equation}
and $\mu$ is a parameter (chemical potential) that can be varied to
fix the desired fraction of removed nodes $q$. The value of $\s_i$ is
related to $h_i$ via the equation: \be \s_i\ =\ {\rm sign}(h_i)\ .
\label{eq:mag}
\ee \\ 

Equations \eqref{eq:Pto_immunization}, \eqref{eq:cavity_fields_zeroT},
\eqref{eq:cavity_messages_zeroT} and \eqref{eq:mag} constitute the
full set of cavity equations of the immunization optimization problem
analogous to optimal percolation since the best inmunizators are those
that optimally destroy the giant connected component.  These equations
can be solved iteratively as follows:
\begin{itemize}
\item Choose a value for $\mu, \gamma, \beta$, and initialize all the state variables $\sigma_i$ and  
$h^{i\to j}$ to random values. 
\item Then iterate Eqs.~\eqref{eq:Pto_immunization}  until 
convergence to find the values of $\nu_{i\to j}$.
\item  Then iterate Eqs.~\eqref{eq:cavity_messages_zeroT} 
until convergence to find the values of $h_{i\to j}$.
\item Compute the new $h^{i}$ using~\eqref{eq:cavity_fields_zeroT}, and the 
the new state $\sigma^i$ of node $i$ via Eq. \eqref{eq:mag}.
\item Repeat until all the fields $\{h_i\}$ have converged.
\end{itemize}

In cases where the equations~\eqref{eq:cavity_messages_zeroT} do not
converge, we use the reinforcement
technique~\cite{braunstein2006learning}.  Once a solution to the
equations have been found, the configuration $\vec{\s^*}$ is the
output of the algorithm: if $\s_i^*=1$ the node is removed, and if
$\s_i^*=-1$ it is present. The $\CI_{\rm BP}$ algorithm has the same
performance as the $\CI_{\rm P}$ algorithm, as we show for the case of
random cubic graphs in Fig.~\ref{fig:rrg_warmald}, reproducing the
exact result of \cite{wormald} for small system size and leaving
virtually no improvement left for these systems. However, while it
improves over CI, it suffers the same deficiency for large systems as
$\CI_{\rm P}$ and BDP since it is a quadratic algorithm which can be
applied only to small networks.

\section{Conclusions}

We have shown how to implement the CI algorithm introduced
in~\cite{MoMa} in nearly linear time when nodes are removed one by
one. This is possible due to the finite radius $\ell$ of the CI
sphere, which in turn allows one to process the CI values in a
max-heap data structure. This trick avoids the full sorting of the CI
values, thus saving exactly a factor $O(N)$.

Moreover, we have introduced $\CI_{\rm P}$, a slightly modified CI
algorithm taking into account the global rearrangement of the CI
values after each node removal, and, in this respect, it corresponds
to the $\ell\to\infty$ limit of CI.  We have also presented $\CI_{\rm
  BP}$, a new algorithm to solve the optimal percolation problem,
which blends the dynamics of the SIR disease spreading model with
message passing updating rules.  The analysis of these algorithms
(including BDP as well) reveals that the improvements over CI are
small and, more importantly, they are made at the expense of
increasing the computational complexity from linear (CI) to quadratic
(BP) in the system size $N$, rendering BP unfit for large datasets.

Therefore, CI remains the viable option of a
nearly-optimal-low-complexity influencer search engine, which is
applicable to massively large networks of several hundred million of
nodes, while the global $\CI_{\rm P}$ algorithm can still be used to
find small corrections in small networks when time performance is not
an issue. Furthermore, from a theoretical point of view, the
simplicity of the CI analysis based on the NB eigenvalue remains as a
good option for theoretical generalization of optimal percolation to
more complicated topologies, as shown in \cite{brain} for brain
network of networks with interdependencies and other more complex
applications that are being presently developed.

\end{document}